 \documentclass{sig-alternate}
\usepackage{array}
 \usepackage{mathtools}
\usepackage{mdwmath}
\usepackage{mdwtab}
\usepackage{eqparbox}
 \usepackage{subfigure}
  \usepackage{multicol}
 \usepackage{float}
 \usepackage{url}
\usepackage{amstext,amssymb,amsmath,epsf}
\usepackage{epsfig}
\usepackage{subfigure}
\usepackage{booktabs} 
\usepackage{color}
\definecolor{lightred}{rgb}{0.8, 0.1, 0.1}
\definecolor{lightblue}{rgb}{0.1, 0.1, 0.8}
\pdfoutput=1
\begin{document}
%

\title{Activity recognition for a smartphone and web based travel survey}



%
%
%
%
%

\numberofauthors{6} 
%
\author{
%
%
\alignauthor
Youngsung Kim \\
       \affaddr{Singapore-MIT Alliance for Research and Technology (SMART)}\\
       \email{youngsung@smart.mit.edu}
\alignauthor
Francisco C. Pereira \\
       \affaddr{Singapore-MIT Alliance for Research and Technology (SMART)}\\
      \email{camara@smart.mit.edu}
\alignauthor
Fang Zhao \\
       \affaddr{Singapore-MIT Alliance for Research and Technology (SMART)}\\
      \email{fang.zhao@smart.mit.edu}
\and  
\alignauthor
Ajinkya Ghorpade\\
       \affaddr{Singapore-MIT Alliance for Research and Technology (SMART)}\\
      \email{ajinkya@smart.mit.edu}
\alignauthor
P. Christopher Zegras\\
       \affaddr{Massachusetts Institute of Technology (MIT)}\\
       \email{czegras@mit.edu}
\alignauthor
Moshe Ben-Akiva \\
       \affaddr{Massachusetts Institute of Technology (MIT)}\\
       \email{mba@mit.edu}
}


\maketitle

\begin{abstract}

In transport modeling and prediction, trip purposes play an important role since mobility choices (e.g. modes, routes, departure times) are made in order to carry out specific activities. Activity based models, which have been gaining popularity in recent years, are built from a large number of observed trips and their purposes. However, data acquired through traditional interview-based travel surveys lack the accuracy and quantity required by such models. Smartphones and interactive web interfaces have emerged as an attractive alternative to conventional travel surveys. A smartphone-based travel survey, Future Mobility Survey (FMS), was developed and field-tested in Singapore and collected travel data from more than 1000 participants for multiple days. To provide a more intelligent interface, inferring the activities of a user at a certain location is a crucial challenge. This paper presents a learning model that infers the most likely activity associated to a certain visited place. The data collected in FMS contain errors or noise due to various reasons, so a robust approach via ensemble learning is used to improve generalization performance. Our model takes advantage of cross-user historical data as well as user-specific information, including socio-demographics. Our empirical results using FMS data demonstrate that the proposed method contributes significantly to our travel survey application.

\end{abstract}



\terms{Algorithms, Design, Human Factors, Experimentation}


\section{Introduction}

Human activity recognition research is useful to interpret mobility related phenomena in a city \cite{Song2010ModellingMobility}. Understanding \emph{why} people go to some places at certain times has beneficial ramifications in many fields such as transportation, internet commerce, urban traffic management, location based services, public health, urban planning, public safety, and so on \cite{May2008nextDMchapter}. Activity based modeling for travel demand is gaining popularity in recent years and it requires a large number of observed trips and their purposes to build. Traditionally, data used in activity based modeling is collected through interview-based travel surveys. Collecting a sufficiently large sample requires an extensive effort. The accuracy of the collected data depends on the memory of the participant, so it is a challenge to capture high resolution activities for days with complex activity patterns. Due to these limitations, researchers are exploring new ways to conduct travel surveys using mobile sensing devices. Smartphones are pervasive devices that nowadays people carry with them everywhere. They are ideal devices for travel and activity information logging. We have developed a smartphone based activity-travel survey system,  Future Mobility Survey (FMS) \cite{Cottrill2013FMS}, and recently used it in large-scale data collection effort in Singapore.

 FMS acquires movement data through sensors (such as GPS, GSM, WiFi, and Accelerometer) commonly available in current smartphones. Besides the hardware sensors, FMS acquires activity and transportation information through a web-based interactive process. The task of the participant is to check that the stop locations, activities, times, and modes are accurately described (and correct them if necessary) on a web interface. To ensure quality of validated data, the user must accurately label the activity at each stop location. Machine learning based approaches for activity recognition can automate some of these tasks, reduce user burden, and therefore assist the user in providing much needed high quality data. Currently a new version of FMS software is being developed based on data acquired during a field-test to create a more intelligent backend and interface.
 In this paper, we present a learning based model for the activity recognition task.

However, prediction of human activity is a nontrivial task, especially in an urban area. One of the reasons is that activities  often have heterogeneous patterns within a small area (e.g. shopping malls with healthcare facility, supermarket, offices) or at the same time (e.g. working at home; shopping while waiting for the train).  Also, sensor data quality itself is not always the best (e.g. GPS unavailable in indoor activities).

To alleviate uncertainty of real world data, we extract heterogeneous features and merge multiple hypothesis models learned from different user populations. The user's likelihood of performing a certain activity at a given location will depend on user's personal needs which will be driven by his/her socio-demographic characteristics \cite{Kwan1999Gender}. Usually environmental context at the given location limits the type of activities
one can perform. We can also derive the activity likelihood from the activities performed by general population apart from individual user characteristics. In this paper, we present a learning model based on spatial, temporal, and contextual features and conduct various experiments to demonstrate its veracity.
The contributions of this paper are:
\begin{itemize}
\item A method to generate a set of predictive features based on location, time, transition context, and environment context (e.g. Points of Interest),
\item Spatial data quantization methods to balance the noise effect in real world data,
\item Improvement of generalization performance by merging of intra-user data and inter-user data including user's social-demographic information,
\item Analysis of number of training days required for a learning model in a real world application.
\end{itemize}

 This paper is organized as follows. In section 2, we review related work. In section 3, we present FMS, a smartphone based activity-travel survey where the proposed activity recognition algorithm will be used. In section 4, we present the proposed activity recognition framework. Extensive experiments are followed with different settings of feature in section 5. Finally, we conclude this paper with some remarks and future work in section 6. \\

 \section{Related work}
With the advance of sensing technology, GPS loggers, and more recently, smartphones, have become popular tools to conduct travel surveys that are essential for transportation planning and management \cite{Bohte2009, Cottrill2013FMS}. The identification of activities is perhaps the most challenging data processing task involved in such travel surveys. The activity categories typically include home, work, social, shopping, pickup/drop-off etc.

Most of the algorithms used to derive activities in GPS travel surveys are rule-based and rely heavily on GIS information, such as Point Of Interest (POI) and land use information \cite{Wolf2001, Huang2010Activitygps, Furletti2013InferringActivities}. An early car-based study in America by Wolf et al. \cite{Wolf2001} inferred trip purposes from GPS data and an extensive GIS land use database.  In more recent work, POI's attractiveness is defined along time of day to indicate the potential possibilities for activities \cite{Huang2010Activitygps}, and \cite{Furletti2013InferringActivities} proposed to infer an activity based on the distance between POI and the stop location. Another option is to use individual characteristics as input for activity recognition algorithms. Axhausen et al \cite{Axhausen2004} developed a rule based approach to identify activities based on users' home and work locations, and POI/land use information in the Swiss. Similar information and rules were used in the GPS survey in the Netherlands \cite{Bohte2009}. Reference \cite{Stopher2008} described a more complicated heuristic rule-based method which collects users' workplace or school, the two most frequently used grocery stores, and occupation beforehand to be used to derive trip characteristics.

More elaborate algorithms have been proposed taking a machine learning approach. Deng and Li \cite{Deng2010} used attributes such as land use, sociodemographic information of the respondents, etc. to construct decision trees. An adaptive boosting technique was used to improve the classification results. Liao et al. \cite{Liao2005Locationbased} proposed a location based activity recognition system using Relational Markov Networks. These works are evaluated based on small samples of experimental data.

Few work exists for activity detection in smartphone based travel surveys. Feldman et al. \cite{Feldman2013iDiary} converted GPS trajectories collected by smartphones into lists of activities by first finding businesses around a user stop, and then employing reverse Latent Semantic Analysis (LSA) to look up the most relevant terms associated with the businesses.


\section{Smartphone-based activity travel survey}
In this section, we give an overview of the FMS system and briefly describe the data which is used for building activity recognition algorithm.

\subsection{Future Mobility Survey (FMS): activity-travel data collection method} \label{sec:FMS}
\begin{figure}[H]
\centering
\includegraphics [width=2.9in]{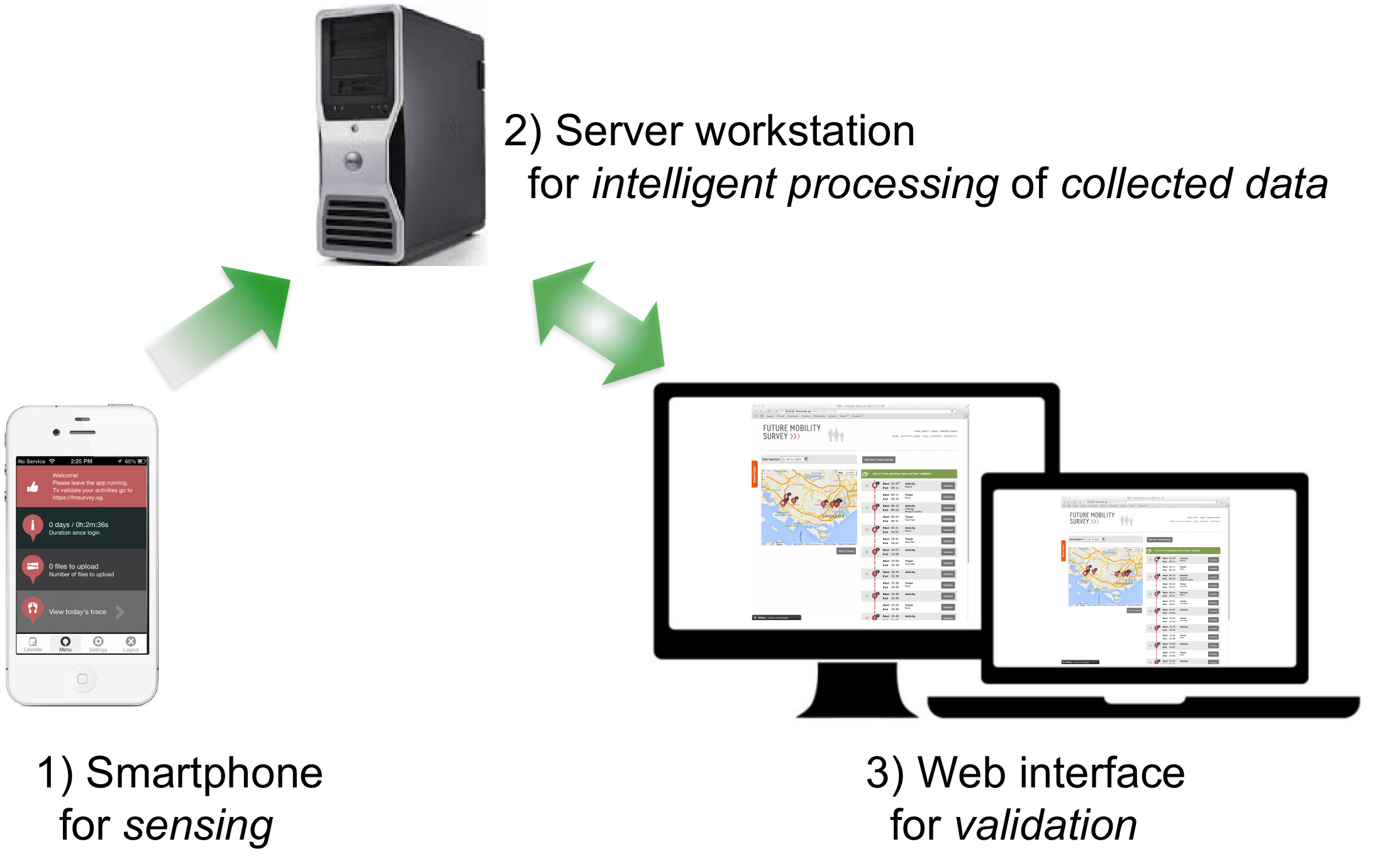}
\caption{Overview of Future Mobility Survey system. The FMS web interface can be found at \textit{{http://www.fmsurvey.sg/}}. }
\label{fig:fms}
\end{figure}

Future Mobility Survey (FMS) \cite{Cottrill2013FMS} collects mobility records through a smartphone application (Android and iOS) and an interactive web interface.  It acquires movement data through sensors commonly available in current smartphones, namely Global Positioning System (GPS), WiFi, Mobile Communications System (GSM, CDMA, and UMTS), and Accelerometer. Stop and mode detection algorithms are run in the backend on the collected raw data and the output is presented to the user in the form of an activity diary \cite{Pereira2013, Cottrill2013FMS}. users can then ``validate" their data by confirming or correcting the system generated stops/modes. In the current FMS system, there is a simple rule-based algorithm to detect only ``home", ``work", and ``change-mode'' activities. The overall flow is depicted in Figure \ref{fig:fms}.

FMS was recently deployed in Singapore \cite{Pereira2013} to conduct a travel survey. Thus far, the FMS has collected collected a total of 22,170 days from 1,440 users in real life situations (more than 130 Million GPS points in total). Among the days and users, we have a total of 7,856 validated days from 948 users. A total of 793 users fully participated in this venture, each one required to collect data for at least 14 days and validate least 5 days. The survey was conducted between October 2012 and September 2013.  Due to battery limitations,  the smartphone application cannot continuously collect the high quality data (e.g. high accuracy GPS and big frequency accelerometer), and as a consequence, the records are sparse in practice. Furthermore, some sensors are not available in certain contexts (e.g. GPS unavailable indoors, WiFi unavailable without nearby APs).



 To our knowledge, FMS is the only smartphone based travel survey that has gone through a field-test with large number of users. Most existing applications \cite{Huang2010Activitygps, Furletti2013InferringActivities, Liao2005Locationbased, Feldman2013iDiary} have used limited size of data collected by fewer than 28 users. The large amount of real world data collected presents a unique opportunity to develop and test machine learning algorithms for activity recognition.

\subsection{Activity categories}
Within the FMS, we have defined seventeen different activities. Home, Work, Work-Related Business, Education, Change Mode/Transfer, Pick Up/Drop Off, Meal/Eating Break, Shopping, Personal Errand/Task, Medical/Dental (Self), Social, To Accompany Someone, Recreation, Entertainment,  Sports/Exercise, Other's Home, and Other. `Other' will be excluded in our activity recognition algorithm.

\section{Methodology}
 In this section, we first present a spatial quantization technique to get empirical activity probability based features. We then describe the ensemble learning based classification methodology using heterogeneous features for different user populations.

\subsection{Spatial-temporal data representation and quantization}

\subsubsection{Data representation}

Our dataset consists of a sequence of $n$ \emph{stop} points for a user $u$, $\{p_i^u | i=1,2,\ldots,n$, and $u = 1,2,\ldots,U \}$, where the user  stayed for a relevant time window\footnote{The FMS minimum threshold is 1 minute to capture mode changes, but it is normally aggregated (by the system or by the user) to much longer chunks.}. Further each stop point is represented as $p_i^u = (x_i, y_i, t_{i1}, t_{i2})$, where $x_i$ and $y_i$ denote the geographical coordinates, $(t_{i1}, t_{i2})$ denotes the start and end time respectively. For simplicity, we use $p_i$ instead of $p_i^u$ now on.

\subsubsection{Data quantization}
 The quantization is applied to the location and time space to enhance data interpretation in terms of context. This context is coarse-grained in spatial and temporal axes. For example, we can deduce a ``transportation change mode" during ``evening rush hour" or deduce that a person may be at ``shopping mall" on ``Sunday evening".  Here, we apply quantization as follows (where $\mapsto$ represents a mapping relationship):

  \begin{itemize}
  \item Spatial cell: the location $(x_i, y_i)$ $\mapsto$ a cell $c_i$.
  Distribution of activities is non-uniform across geographies. Dependent on a mapping function, samples in a cell are different. Some spatial quantization methods will be proposed in section \ref{sec:spatialquantization}.

  \item Set of time slots (within the day): the time period $(t_{i1}, t_{i2})$ $\mapsto$ a set  $\mathcal{S}_i$ of time slots (e.g. 10 minute slots). For example, an activity started at 8:53 and ending at 9:08 will be assigned to a time slot set  $\mathcal{S}$=\{8:50, 9:00, 9:10\}. This works as an ``temporal alignment" step that will later be useful for calculating temporal frequency features.
\end{itemize}

Hence, our dataset will consist of activity points $q_i$ (the quantized version of $p_i$), defined as the tuple $(c_i, \mathcal{S}_i, a_i )$ where $a_i$ denotes an activity from the set of sixteen categories mentioned above. We also create two useful functions: $\mathcal{W}(s)$ returns the day type of a time slot $s$ (weekend or weekday); $\mathcal{X}(c)$ retrieves the set of Points of Interest from our database, corresponding to cell $c$.

 \subsubsection{Spatial quantization methods (distribution adaptive quantization) } \label{sec:spatialquantization}
As mentioned above, the function mapping the location of $p_i$ to a cell $c_i$ affects the likeliness of activity $a_i$ so we explore different mapping (spatial quantization) functions to find an appropriate population representation.
The simplest and easiest way is to divide space arbitrarily regardless of a sample distribution. An adaptive way is to apply the data distribution. In this work, we consider both fixed quantization and dynamic quantization. In the fixed case, once space of training data is quantized, it is used in future probability calculations. In the dynamic case, space is divided when a new instance is identified. In this case, if there are $N$ samples to calculate frequencies, the number of cells is $N$.

\paragraph{Fixed cell}
\begin{itemize}
\item Rectangle shape: quantization is not correlated with regional distribution. The easiest way is to adopt a rectangle shape; parameters including width (horizontal) and height (vertical) size.
\item Voronoi tessellation based polygon: spatial data clusters can be found to apply regional characteristics. Based on a centroid of each cluster, edges and vertices of each cell can be found by Voronoi tessellation. To find an appropriate cluster is a essential process.

\end{itemize}

\paragraph{Dynamic (instance based) cell}
\begin{itemize}
\item Circular polygon: a cell is defined within predefined distance (radius of circle) at each instance. Every instance is a centroid of a cell.
\end{itemize}

 \subsection{Proposed features}

\subsubsection{Activity Frequency}

For each activity point $q_i$, we determine three kinds of activity frequency: Temporal activity frequency, Spatial activity frequency, and Contextual activity frequency. We essentially make use of the following general empirical conditional probability distribution (we use the kronecker delta notation, where $\delta_{i,j}=1$ if $i=j$, and $0$ otherwise):
%

\begin{equation}
\label{eq:prob2}
Pr( a_i = l | b_i) \coloneqq \frac{\sum_{j=1}^N \delta_{a_j,l} \cdot \delta_{b_j,b_i}  }{\sum_{l=1}^L \sum_{j=1}^N \delta_{a_j,l} \cdot \delta_{b_j,b_i} }
\end{equation} where $N$ denotes the total number of activity points in the same cell for all users $u \in \mathcal{U}$ ($\mathcal{U}$ is a user set), $b_i$ denotes a bin, and $l$ denotes an activity type ($L$ is total number of activities).

\noindent In this equation, we count a normalized frequency of activity $l$, within a bin over the total count of all activities within the same bin. For spatial activity frequency, the bin we use is a spatial cell $c_i$.

In order to estimate the temporal activity frequency, we need a slightly more sophisticated treatment of the data. In this case, the statistics depend on the time slot sequence of the activity points, where each time slot adds 1 (e.g. an activity that spans from 8:00 to 10:00 contributes 12 to the total count, assuming 10 minutes time slots). The bin at activity point $i$ is now defined by its entire sequence of time slots ($\mathcal{S}_i$). Inclusion or exclusion of a different activity point $j$ in that bin is based on how many common time slots exist between $i$ and $j$.

For the contextual activity frequency, we first map each POI category to one of the sixteen activity classes and then compute a relative frequency of each activity type in each spatial cell.

\begin{figure}[tb]
 \centering 
 \begin{tabular}{c c}
 \hspace{-0.2in}\subfigure[Rectangle cell]{\includegraphics[width=1.5in]{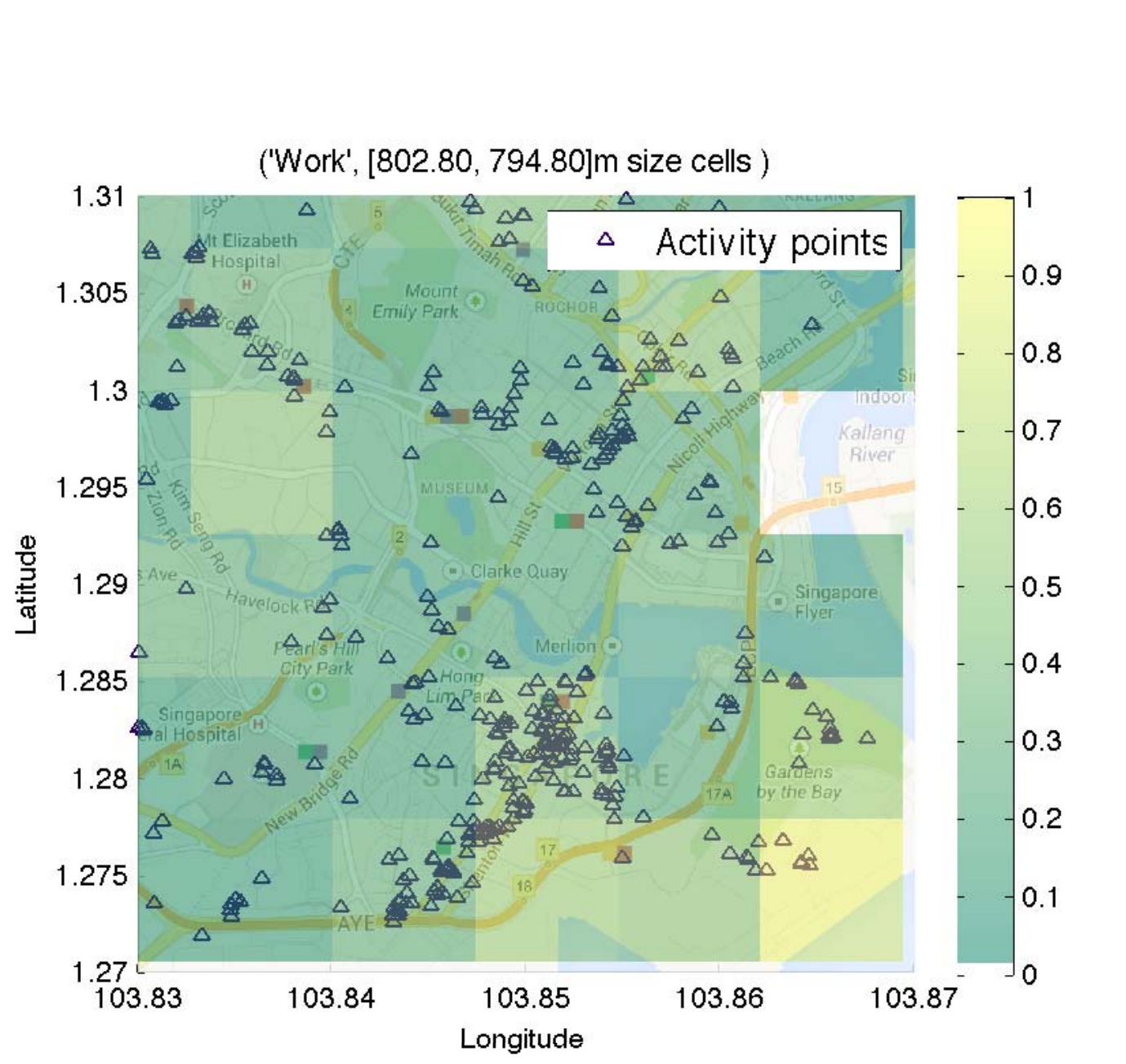} }&%
 \hspace{-0.2in}\subfigure[Polygon cell]{\includegraphics[width=1.5in]{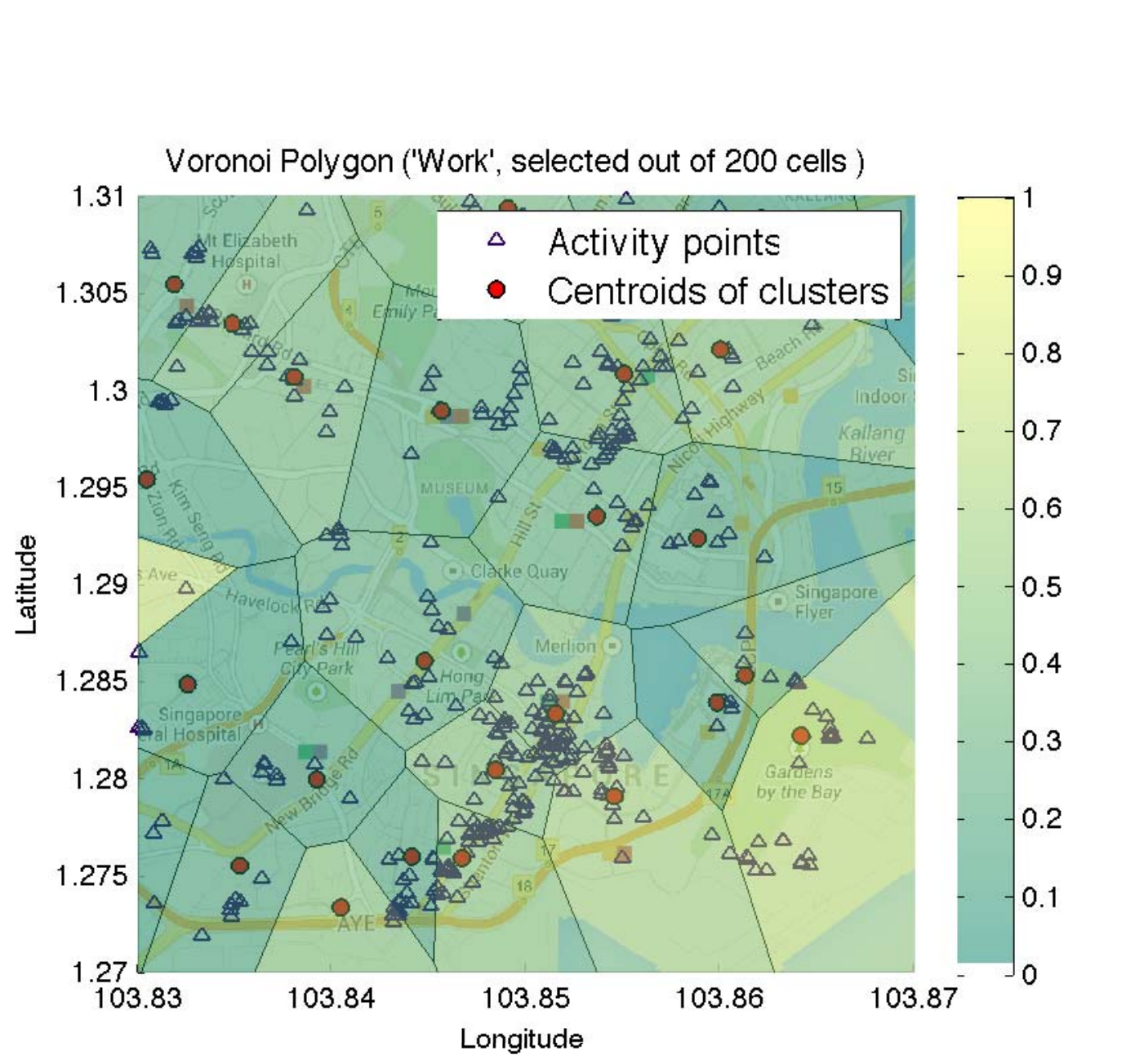}}\\%
 \hspace{-0.2in}\subfigure[Circular Polygon Cell]{\includegraphics[width=1.5in]{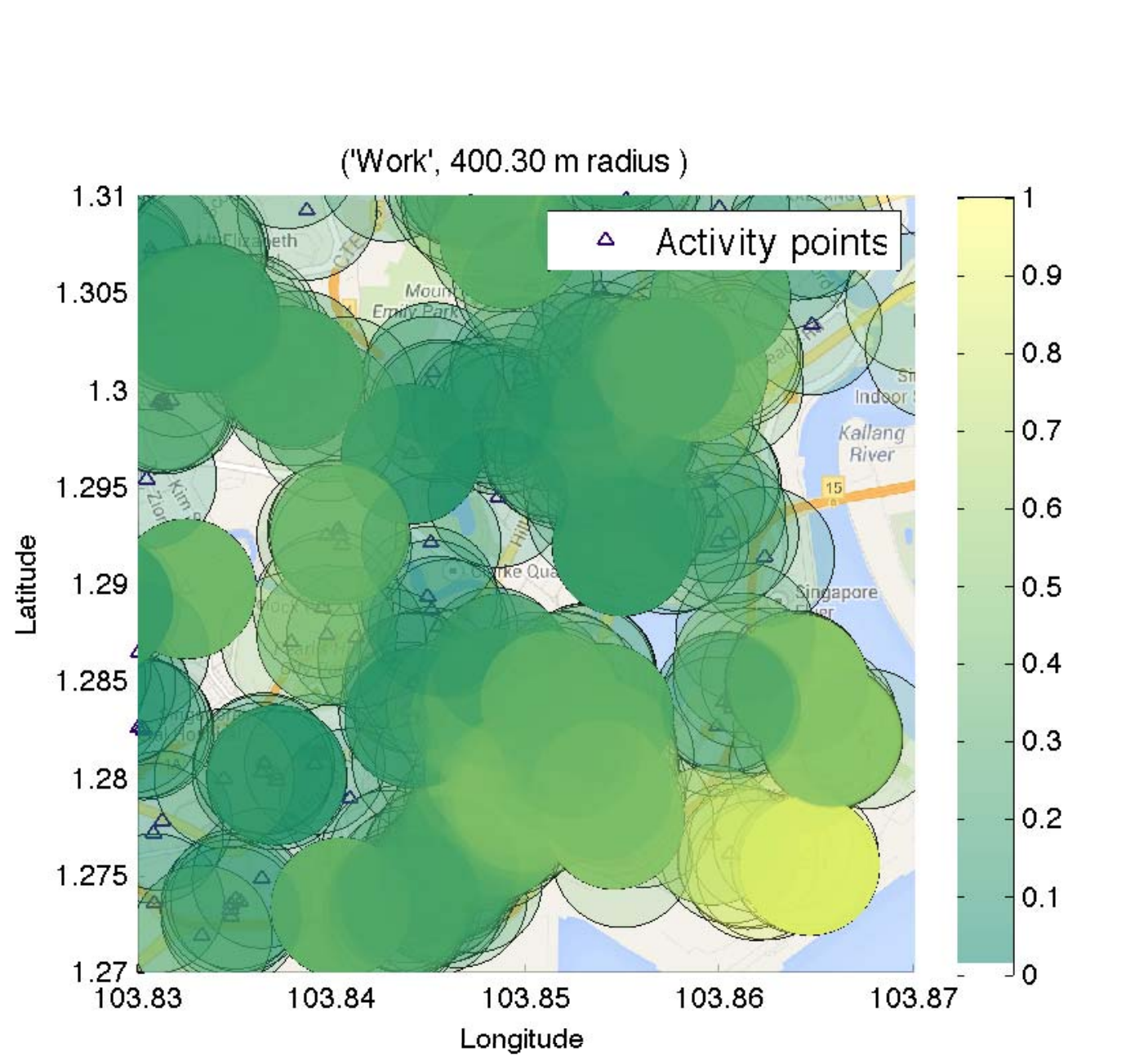}} &
 \hspace{-0.2in}\subfigure[Skeleton]{\includegraphics[width=1.2in]{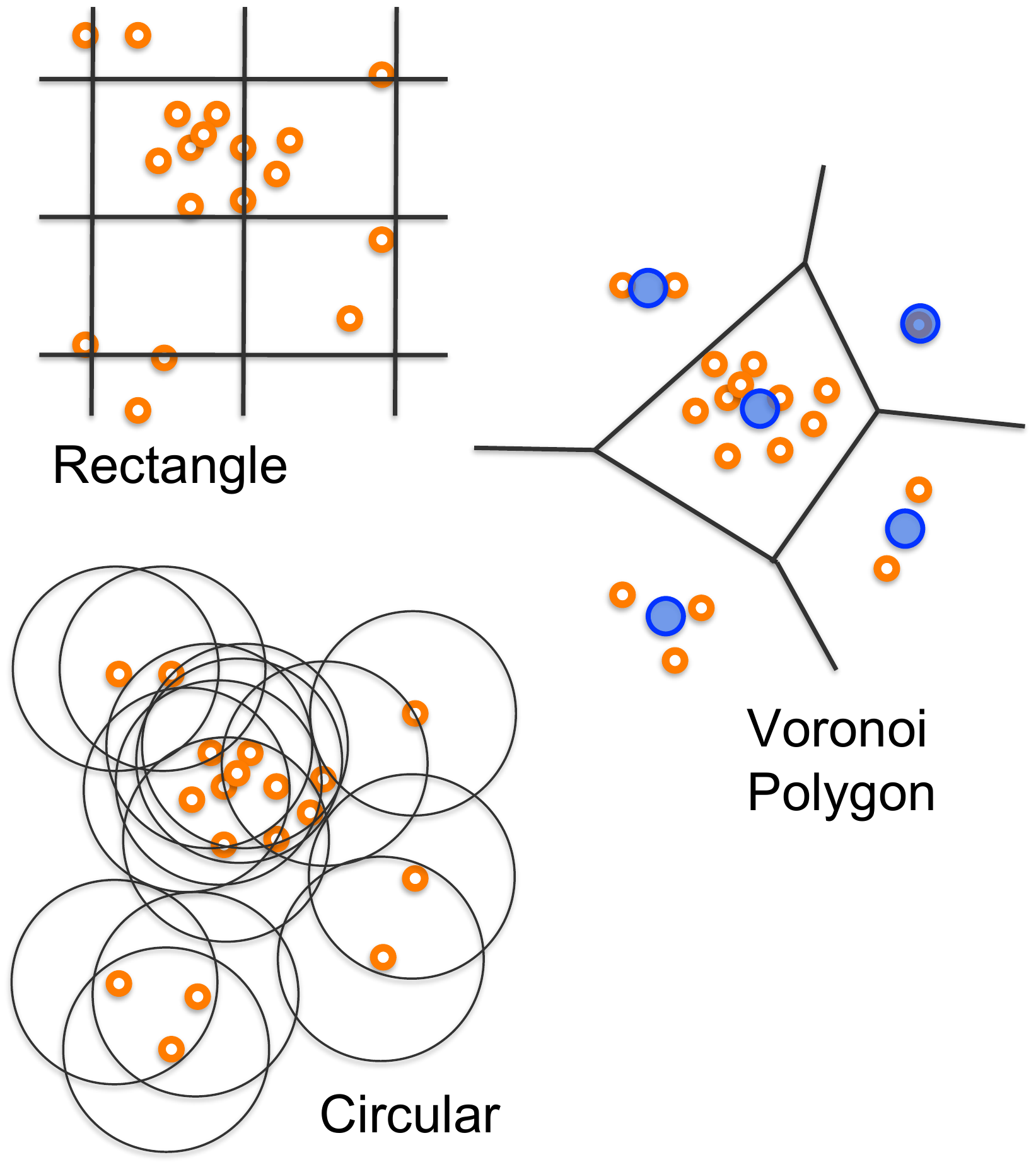}} \\%
\end{tabular}
\caption{\textbf{Empirical probability $p(a_i|c_i)$ in (\ref{eq:prob2}) of \textsf{Work} activity in spatial cells} are shown: (a) rectangle cells, (b) polygon cells centroids of clusters, (c) circular cells at each activity point. (d) simplified explanation.}
\label{fig:celltypes}
\end{figure}

In Figure \ref{fig:celltypes} (a), (b), and (c), the spatial activity frequency as calculated through equation (\ref{eq:prob2}) is depicted using real data for different cell types defined in the previous section. Colormap indicates a degree of the probability.

\subsubsection{Distance based empirical probability}
For each point $p_i$, we obtain distance related features using Euclidean distance. We define the distance between a point $p_i$ and a set of points $P$ as $d(p_i, p_j)=min\{{\|p_i - p_j\|}_2 : p_j \in P\}$. These features are calculated with respect to POIs, past activity information from all users and home and work from the specific user. Firstly, for cell $c_i$ containing the point $p_i$ we obtain the contextual neighbor activity confidence 
\begin{equation}\label{eq:prob3}
Pr(a_i=l|\mathcal{X}_l(c_i))  \coloneqq    \phi(d(p_i,\mathcal{X}_l(c_i)))
\end{equation}
and the historical neighbor activity confidence
\begin{equation}\label{eq:prob4}
Pr(a_i=l|\mathcal{A}_l(c_i))  \coloneqq    \phi(d(p_i,\mathcal{A}_l(c_i)))
\end{equation}
where $\mathcal{X}_l(c_i)$ is the activity set of type $l$ from contextual data (POIs) in cell $c_i$, $\mathcal{A}_l(c_i)$ is the activity set of type $l$ from  in cell $c_i$, and $\phi(\cdot)$ can be any activation function such that it is normalized between 0 and 1. We have used $\phi(d)={(1+d^2)}^{-1}$ which is the inverse of the squared distance. (Also, a distance $d$ is normalized between 0 and 1 for the points in the same cell).

Secondly, for  each user $u$, we choose ``core" activities (home and work), and calculate their core activity distance to $p_i$.

\subsubsection{Activity Transition Probability} For each point $p_i$, we obtain activity probability based on the previous activity. The simplest way is to apply the first-order Markov chain where a current activity ($a(t)$) is conditioned on the value of most recent previous activity ($a(t-1)$) in a transition distribution. We calculate the empirical transition probability:

  \begin{eqnarray}\label{eq:transition}
Pr_{sl}(t-1,t) =  &  Pr(a(t) = l | a(t-1) = s)   \nonumber  \\
      \coloneqq   &   \frac{\sum_{j=1}^{N}\delta_{a_j(t), l}\cdot \delta_{a_j(t-1), s} } {\sum_{l=1}^L\sum_{j=1}^{N}\delta_{a_j(t), l}\cdot \delta_{a_j(t-1), s}} ,   
\end{eqnarray}
where $N$ denotes the total number of activity points for all users $u \in \mathcal{U}$, $l$ denotes the current activity, $s$ denotes the previous activity, $l,s \in \mathcal{A}$ ($\mathcal{A}$ is an activity set), and  $\sum_{l=1}^L Pr_{sl} =1$.

\noindent  We apply equation (\ref{eq:transition}) to historical data to obtain transition probability matrix. Due to varying patterns during weekends and weekdays, we obtain two transition matrices for corresponding periods. In practice, if there is no previous activity (no activity reported within 24 hours), we assume a uniform probability for each activity. We use these probability matrices to calculate the activity probability of current point $p_i$.

\subsubsection{Activity duration} For each point $p_i$, we calculate its activity duration, $T_i=(t_{i2}-t_{i1})$. \\

Acceleration and speed features are excluded since activity defined here is not about physical behavior such as walking, running, and so on \cite{Kwapisz2011ActivityAccelerometers}. These features are used to detect stop segments in the FMS system as mentioned above. \\

After the feature extraction process explained above we have the following feature vector, general features
\begin{equation}
\begin{split}
\mathbf{x} =  & [ \text{Temporal Activity Probability} \in \mathbb{R}^{1\times L}, \\
                      & \text{Spatial Activity Probability} \in \mathbb{R}^{1\times L}, \\
              	     & \text{Contextual Activity Probability} \in \mathbb{R}^{1\times L}, \\
          	     & \text{Activity Transition Probability} \in \mathbb{R}^{1\times L}, \\
                     & \text{Historical Neighbor Activity Confidence} \in \mathbb{R}^{1\times L}, \\
                     & \text{Contextual Neighbor Activity Confidence} \in \mathbb{R}^{1\times L}, \\
                    & \text{Core Activity Distances} \in \mathbb{R}^{1\times 2}, \\
                    &  \text{Activity Duration} \in \mathbb{R}^{1} ]^T \in \mathbb{R}^{6L+ 3},
\end{split}
 \end{equation}
 where $L$ is the number of activity categories.\par

\subsection{Classification}
 When the data is acquired from multiple sensors or sources (and then heterogeneous features are generated), a single classifier cannot find good decision boundary for classification \cite{Polikar2006Ensemble}. To overcome this problem, in this section, we present ensemble learning based classification. Ensemble learning, here, is used through two levels; one is to learn heterogeneous features, and in second step, outputs from classifiers such as score and decision are merged to a final decision.

\subsubsection{Ensemble decision trees}
Ensemble learning has been widely used to cope with noisy real world data. In this paradigm, several (base) classifiers are learned from training data to eventually become a unified classifier. In theory, individual base classifiers can concentrate on different areas of the problem space and, as a result, the unified classifier, which combines the output of those base models, becomes more robust. Two kinds of ensemble learning are used in this paper, namely bootstrap aggregating (Bagging) and random subspace. In Bagging, each \textit{base classifier} is trained with a subset generated by subsampling on the global training set. In the random subspace approach, each \textit{base classifier} is learned using subspace features of the original feature set. To predict a class label for unseen data, a majority voting process is applied on the set of individual predictions.

 Our base classifier will be decision trees, one of the popular methods, which consist of gradually splitting the input feature space into decision regions. This method is useful to deal with irrelevant variables and is robust to outliers. However, decision trees show unstable performance. To alleviate instability, ensemble learning has been widely adopted. One popular method is bagging of decision trees. Another powerful tool is a combination of aggregating set of random features (subspace) based on decision tree classifier, namely Random Forests  \cite{Ho1998RandomSubspaceForests}.

Using a set of training features and activity labels $\{\mathbf{x}_i, {a}_i \}  \in \mathnormal{Tr}, \forall i$ where  $\mathnormal{Tr}$ is a training set, we calculate an ensemble hypothesis function $h(\mathbf{x}, \boldsymbol{\Theta} )$ where $\boldsymbol{\Theta}$ is a set of decision tree hypothesis $\theta_k$, $\forall k$. This function finds an activity label ${a}$, based on $a = {\arg \max}_{\substack{l}}~{s_l}$, where $s_l$ is the score for activity label $l$. This function will be used to predict a label of unseen data $\mathbf{x}_{test} \in \mathnormal{Te}$ for test in future.

\subsubsection{Ensemble of user social demographic characteristics based learning}
Users with different social demographic characteristics show different activity and travel patterns \cite{Jiang2012Clustering,Kwan1999Gender}. It is, thus, helpful to learn a model using individual user's history data, in addition to learning from other users' history data. An individual user belongs to multiple categories; formally each user is included in several different user sets: $u \in \mathcal{U}, \mathcal{P}, \mathcal{O}, \mathcal{G}$, where $\mathcal{U}$ denotes a cross (universal) user set, $\mathcal{P}$ denotes a specific user set, $\mathcal{O}$ denotes an age-specific user set, and $\mathcal{G}$ denotes a gender-specific user set. The input feature vector of $p_i$ for a user $u$, $\mathbf{x}({p_i^u})$, (where $u \in {\mathcal{U}}$ and $\mathbf{x}_{\mathcal{U}}, \forall u$), is generated based on subsets. Classifiers (hypotheses) are learned using user subsets: $h(\mathbf{x}_{\mathcal{U}})$, $h(\mathbf{x}_{\mathcal{P}})$, $h(\mathbf{x}_{\mathcal{O}})$, and $h(\mathbf{x}_{\mathcal{G}})$. From each model, we get outputs such as 1) a score vector with a element $s_l \in [0, 1], \forall l$ for each class (activity) label and 2) a decision $d_{l}, \forall l$ for the $l$-th class. The score of each activity class from the hypothesis $h(\cdot)$ become an input feature vector for ensemble classifier to determine a final score. Classifier's decision can be merged by classifier learning and Weighted Majority Voting (WMV). WMV is one popular method to merge multiple decisions to obtain a final decision (based on $\arg\max_l  \sum_{t=1}^T w_{t}d_{t,l}~\forall l$ where  $w_{t}$ is a weight for $t$-th classifier's decision $d_{t,l} \in \{0,1\}$ for $l$-th class.) \cite{Polikar2006Ensemble}.

\subsection{Workflows of the proposed algorithm}
\begin{figure*}[tb]
        \centering
      \includegraphics[width=4.8in]{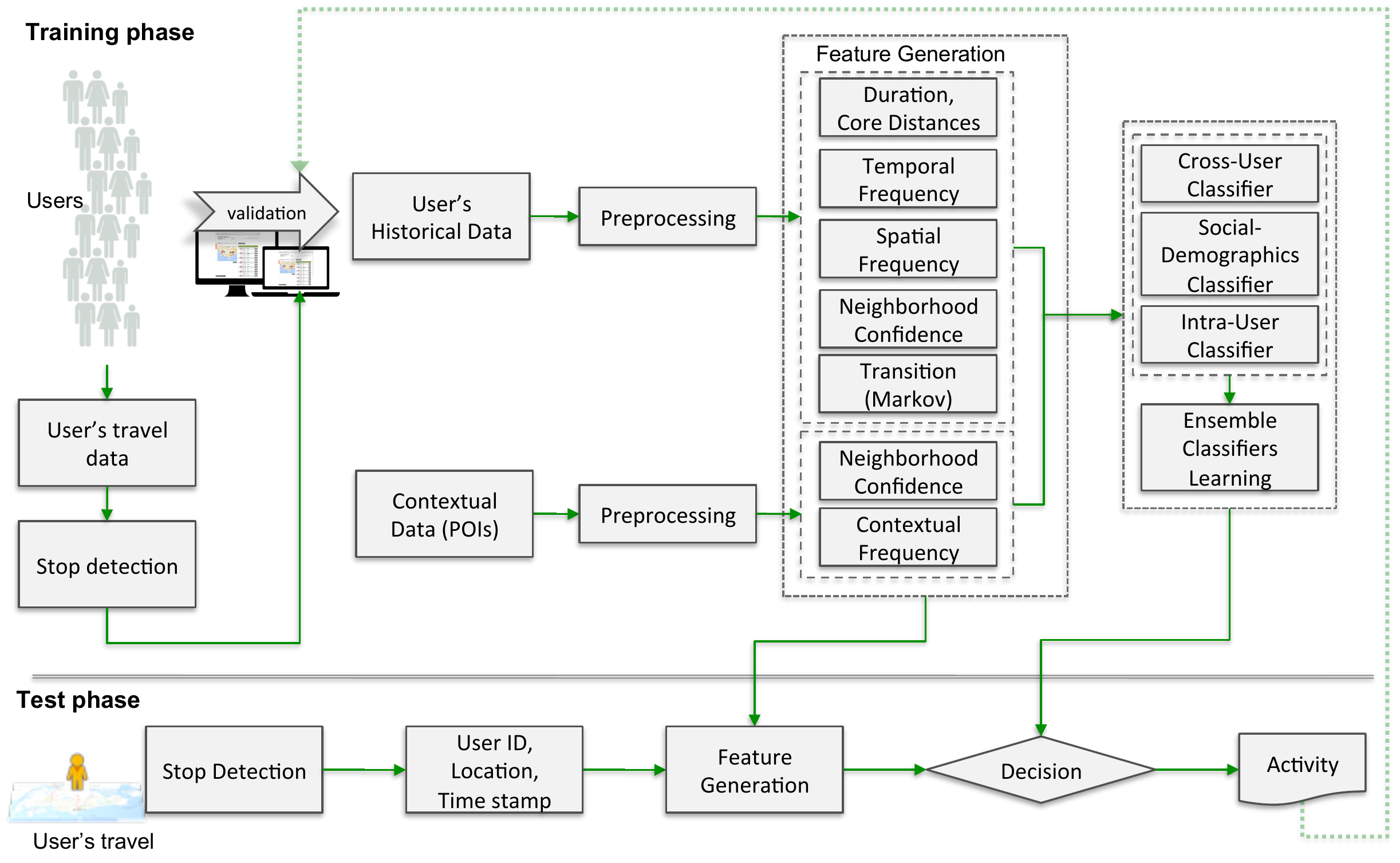}
        \caption{Overview of the proposed activity recognition system. Based on given an identified stop (detected by the current stop detection algorithm), the algorithm identifies an activity based on spatial, temporal, transition, and contextual features. We assume that his/her home location is known beforehand (provided when he/she registered in the website).    }
        \label{fig:overview}
\end{figure*}

Figure \ref{fig:overview} shows an overall flow of the proposed activity recognition system used in FMS. We infer an activity type for each user stop point.

\section{Experiments}
In this section, we evaluate the proposed algorithm using a dataset acquired through our FMS system.

\subsection{Data set}
Within the FMS, we have 793 users who have completed the survey with at least 5 validated days, as mentioned in Section \ref{sec:FMS}. POI data has been provided by Singapore Land Authority (SLA). It has a total of $64,819$ points related to shopping malls, clinics, bus stops, and metro train stations, residential buildings, office buildings and so on\footnote{\scriptsize{POI label includes;e.g. Pub/Bar, Restaurant, Kiosk/Stall, Cafe, Pet Shops, Child Care, Skin Care, Gym, Supermarkets, Convenience Stores, ATMs, MRT Stations, Swimming Complexes, Tuition Centres, Music Dance Schools, Car Wash, Toy Stores, Photography, Post Offices, Town Councils, HDB Branch Offices, Police Stations, Primary Schools, Secondary Schools, Hair Salons, Yoga Pilates, Accountants, Maid Agencies, Clinics, Laundry, Travel Agencies, Religious, Pharmacies, and so on.}}. These POIs are mapped to our 16 activity categories. Table \ref{table:poi} shows the statistics for the mapping.

\begin{table}[htb]
\scriptsize
\caption{The number of environmental context data per activity category generated based on Points of interest (POIs) which contain location information.  } \label{table:poi}
\centering
\begin{tabular}{c c c}
\toprule
Activity & $\#$points & percent ($\%$) \\
\midrule
Home & 31 & 0.05 \\
Work & 48 & 0.08 \\
Change Mode/Transfer & 4965 & 8.25 \\
Pick Up/Drop Off & 0 & 0.00 \\
Shopping & 19862 & 32.99 \\
Social & 0 & 0.00 \\
Work-Related Business & 4619 & 7.67 \\
Education & 2678 & 4.45 \\
Recreation & 888 & 1.48 \\
Medical/Dental (Self) & 4150 & 6.89 \\
Meal/Eating Break & 10200 & 16.94 \\
Entertainment & 181 & 0.30 \\
Sports/Exercise & 529 & 0.88 \\
Personal Errand/Task & 12046 & 20.01 \\
To Accompany Someone & 0 & 0.00 \\
Other's Home & 0 & 0.00 \\
Other &	4670 & -\\
\bottomrule
\multicolumn{2}{l}{*Other is excluded.   }\\

\end{tabular}
\end{table}

\subsection{Data preprocessing and cleaning}
As with any kind of survey, the data collected in FMS contains noise/errors, and this problem may be more serious in this case than average. Since the FMS users were not guided by interviewer in their validation process, the task has been proven to be challenging to some of the users, especially those less tech-savvy users. As a result, there can be multiple errors in user's data. Therefore, data cleaning is an essential step before we perform any performance evaluation. Firstly, we select days where users started and finished their daily activity at home. Then, we apply a sequence of checks, and discard the data if home to home distance is higher than 50 meters; if home to other validated activities is less than 10 meters; or if activity points have swapped time between start and end of one activity. We also apply other filters: no activity with more than 24 hour duration is allowed; an activity outside of Singapore area is removed. As a result, we use 5,073 points from 243 users where their data had been collected from March 11th of 2013 to September 30th of 2013  for the following experiments.

\subsection{Protocols and parameter settings} \label{paramsettings}
First, we apply two-fold validation where we keep the chronological order of data with $k$ training days and one test day split, $k=1,2,3,4$ for every users. In the experiments, we apply different parameter settings: different resolutions of time slot: [10, 20, 40, 60, 90, 120] minutes; different resolutions of spatial cell width: [200, 400, 600, 800, 1000] meters; number of clusters for Voronoi polygons: [1000, 800, 600, 400, 200, 100]; Circle radii: [100, 150, 200, 300, 400, 500] meters.

For the random subspaces based decision trees (Random Forest (RF)), a dimension of subspace features is chosen based on square root of the total number of feature variables. For decision tree-based (DT) classifiers including RF and bagging of DT (BagDT), the minimum number of observations per tree leaf is set as 1. 100 base classifiers are used. A random seed found by pseudorandom number generation is fixed.



\subsection{Results}

\subsubsection{Different resolutions of temporal slot and spatial cell}
Ensemble methods (BagDT and RF) show constant average accuracy as temporal cell size increases. Accuracy value of those methods increases as spatial cell size increases for Rectangle and Voronoi Polygon cases. For more details, a reader can refer to \cite{Kim2014ICPR}.

\subsubsection{Different number of training days}
\begin{figure}[ htb]
 \centering 
 \begin{tabular}{c c }
  \hspace{-0.25in}   \subfigure[Individual Models]{\includegraphics[width=1.9in]{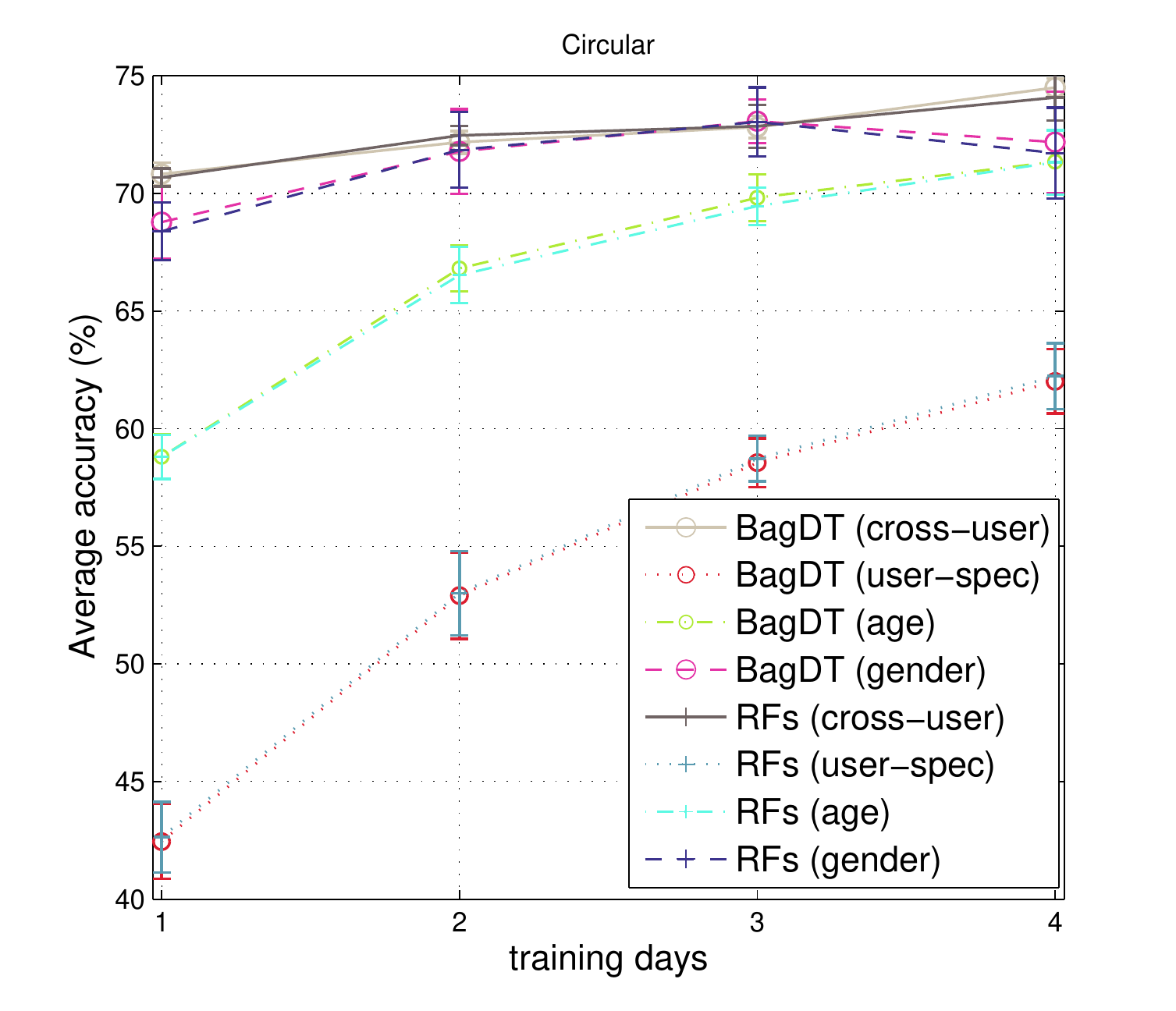} } &
 \hspace{-0.4in}   \subfigure[Fusion Models]{\includegraphics[width=1.9in]{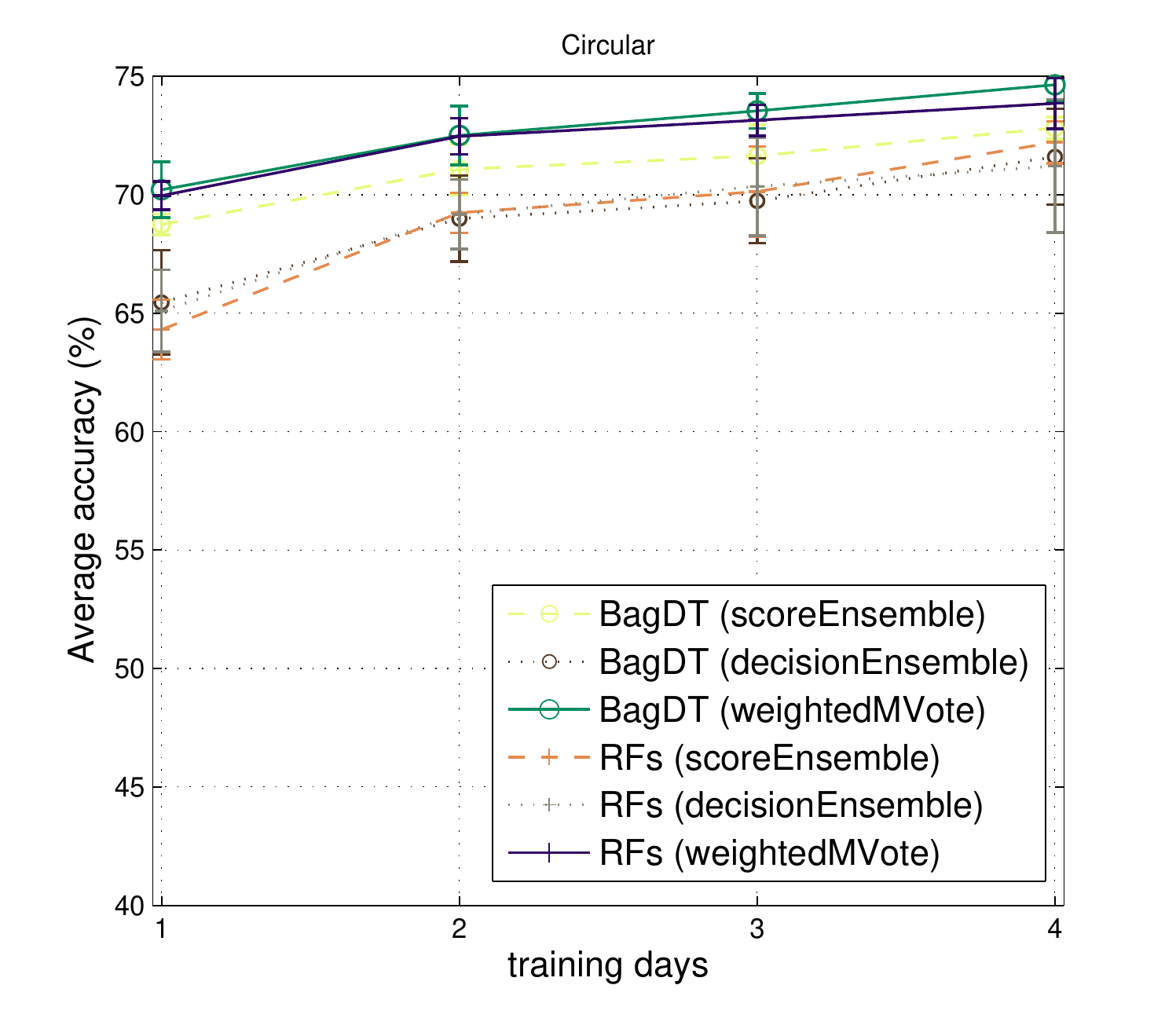}} 
\end{tabular}
\caption{Average prediction accuracy along with number of training days for each model: (a) individual classifier learned using different sets of user population.  (b) ensemble classifiers for merging individual classifier models. }
\label{fig:trainingdays}
\end{figure}

Figure \ref{fig:trainingdays} shows the average classification accuracy for different number of training days. We see that the average accuracy is improved as the number of training days increases. In Figure \ref{fig:trainingdays} (a), individual classifier was learned using different sets of user population such as cross-user, individual user, age-specific user, and gender-specific user. The model using more training data shows better classification performance. Due to small number of training samples, user-specific model solely does not show best performance. However, that accuracy value drastically increases compared to other models as data size increases. In Figure \ref{fig:trainingdays} (b), classification performance of ensemble of individual models are shown. Ensemble models show better classification performance than that of individual models. Decision fusion based on weighted majority voting (weightedMvote) methods show stable and best performance along with the training days as shown in Figure \ref{fig:trainingdays} (b).

%
%
%

\subsubsection{Relationship between activities and merging}
In Table \ref{table:confusion}, we show classification confusion matrix for 16 activity categories. As shown in the table,
most of the points in the Pick Up/Drop off class (PD) is classified as Change Mode/Transfer (C). Work-Related Business (WR) activities are mainly classified as Work (W). Many other activities (related to maintenance or discretionary context) are classified as Change Mode/Transfer (C) which has the largest training sample size. And this may relate to the fact that many shopping malls and shops are located close to street and bus/train stations in Singapore.

As the 16 activities cannot be exclusively explained, i.e. more than one activity can be tagged for one certain user stop point. We follow the work of  \cite{Kulkarni2000Microsimulation, Recker1985Travel} to distill this set into a set of conceptually exclusive activities: 1) Home, 2) Work (including \textit{Work}, \textit{Work-Related Business}, and \textit{Education}), 3) Transportation (including \textit{Change Mode/Transfer} and \textit{Pick Up/Drop Off}, and 4) Maintenance/Discretionary (including \textit{Meal/Eating Break}, \textit{Shopping}, \textit{Personal Errand/Task},  \textit{Medical/Dental (Self)}, and so on). Table \ref{table:fusionlevel16and4} shows that classification accuracy using four activity definition is improved compared to full sixteen activity categories.

\begin{table*}[ htb]
\tiny
\caption{Confusion matrix: Random Forests (RF) prediction of Table \ref{table:freq_type} }\label{table:confusion}
\centering
 \begin{tabular}{c| c c c c c c c c c c c c c c c c c }
\toprule
truth $\backslash$predict  & H   & W  & C & PD & Sh &So & WR & E & R & MD & M & E & Sp & P & A & OH   &accuracy ($\%$)  \\
\hline
 H   &\underline{63}	&0	&0&	0&	0&	0&	0&	0&	0&	0&	0	&0	&0	&0	&0	&0	&100\\
W	&3	&\underline{105}&10&	0&	0&	0&	0&	0&	0&	0	&1	&0	&0	&0	&0	&0	&88.24\\
C	&0	&1	&\underline{147}&	0&	0&	0&	1&	0&	0&	0&	1	&0	&0	&2	&0	&0	& 96.71\\
PD	&0	&0	&3&	\underline{1}&	0&	0&	0&	0&	0&	0&	1	&0	&0	&1	&0	&0	& 16.67\\
Sh	&0	&1	&7&	0&	\underline{1}&	0&	1&	0&	0&	0&	3	&0	&0	&0	&0	&0	& 7.69\\
So	&0	&2	&7&	0&	0&	\underline{0}&	0&	0&	0&	0&	2	&0	&0	&0	&1	&0	&0\\
WR	&0	&8	&5&	0&	0&	0&	\underline{0}&	0&	0&	0&	1	&0	&0	&0	&0	&0	&0\\
E	&1	&4	&3&	0&	0&	0&	0&	\underline{1}&	0&	0&	1	&0	&0	&0	&0	&0	& 10.00\\
R	&0	&0	&1&	0&	0&	0&	0&	0&	\underline{0}&	0&	0	&0	&0	&0	&0	&0	&0\\
MD	&0	&0	&0&	0&	0&	1&	0&	0&	0&	\underline{0}&	1	&0	&0	&0	&0	&0	&0\\
M	&0	&5	&13	&0	&1	&0	&2	&0	&0	&0	&\underline{29}	&0	&0	&1	&0	&0	& 56.86\\
E	&0	&0	&0&	0&	0&	0&	0&	0&	0&	0&	1	&\underline{0}	&0	&0	&0	&0	&0\\
Sp	&0	&0	&0&	0&	0&	1&	1&	0&	0&	0&	0	&0	&\underline{0}	&0	&0	&0	&0\\
P	&0	&2	&7&	0&	0&	0&	2&	0&	0&	0&	0	&0	&0	&\underline{4}	&2	&0	& 23.53\\
A	&0	&1	&1&	0&	0&	0&	0&	0&	0&	0&	0	&0	&0	&0	&\underline{1}	&0	& 33.33\\
OH	&0	&0	&0&	0&	0&	0&	0&	0&	0&	0&	0	&0	&0	&0	&0	&\underline{0}	& - \\
\hline
Overall & & & & & & & & & & & & & & & & &    75.54 \\
\bottomrule
\multicolumn{18}{l}{Home (H), Work (W), Change Mode/Transfer (C), Pick Up/Drop Off (PD), Shopping (Sh), Social (So),  }\\
\multicolumn{18}{l}{Work-Related Business (WR), Education (E), Recreation (R), Medical/Dental (MD), Meal/Eating Break (M), }\\
\multicolumn{18}{l}{Entertainment (E), Sports/Exercise (Sp), Personal Errand/Task (P), To Accompany Someone (A), Other's Home (OH)}\\
 \end{tabular}\label{table:confusion}
\end{table*}

\subsubsection{Prediction performance improvement by merging of different sets of user population }
Table \ref{table:fusionlevel16and4} shows classification accuracy for 16 classes and 4 classes respectively, using 4 training days and Random Forest with Rectangle cell type. Scores of multiple classifier learned using different user population is merged by a classifier (Scores Ensemble by classifier). Decisions from multiple classifiers are merged by classifier (Decisions ensemble by classifier) and by Weighted Majority Voting. For the weighted majority voting (WMV), weights are simply determined with `4' for the cross-user model, `3' for the gender model, `2' for the age model, and `1' for intra-user model. This is based on number of training samples per model; General (total) > Gender > Age > User-specific. Decision merging with WMV shows consistently better classification accuracy than to other models.
 	
 \begin{table}[ htb]
 \small
\caption{Overall accuracy (number correctly classified/total number of samples), Random Forests}\label{table:freq_type}
\label{table_result}
\centering
\begin{tabular}{l c   }
\toprule
method  & accuracy($\%$)    \\
\hline
 \multicolumn{2}{c}{16 classes }\\
\hline
cross-users 				 	&  72.32 \\
userID			 		 	&  63.95  \\
Age						 	&  70.60 \\
Gender				 	 	&  74.68 \\
Scores ensemble (classifier)		 	&  73.61 \\
Decisions ensemble (classifier)			& 73.18 \\
Decisions ensemble	(weighted majority)   &  75.54 \\
\hline
 \multicolumn{2}{c}{4 classes }\\
\hline
cross-users  					& 78.98\\
userID					        & 74.95\\
Age							& 80.89\\
Gender						& 83.23\\
Scores ensemble (classifier)		& 84.50\\
Decisions ensemble (classifier)		& 83.44\\
Decisions ensemble (weighted majority)         & 84.08\\
\hline
 \bottomrule
 \multicolumn{2}{l}{*setting: 4 training days, 800m $\times$ 800m rectangle size,  }\\
  \multicolumn{2}{l}{120 mins time slot.  }\\
  \multicolumn{2}{l}{*4 Classes: 1) \textsf{Home}, 2) \textsf{Work}, 3) \textsf{Transportation},  }\\
  \multicolumn{2}{l}{ 4) \textsf{Maintenance/Discretionary}}\\
\end{tabular}\label{table:fusionlevel16and4}
\end{table}

\subsubsection{Testing on real data stream and unseen user effect}
\begin{figure}[htb]
\centering
   \epsfig{file=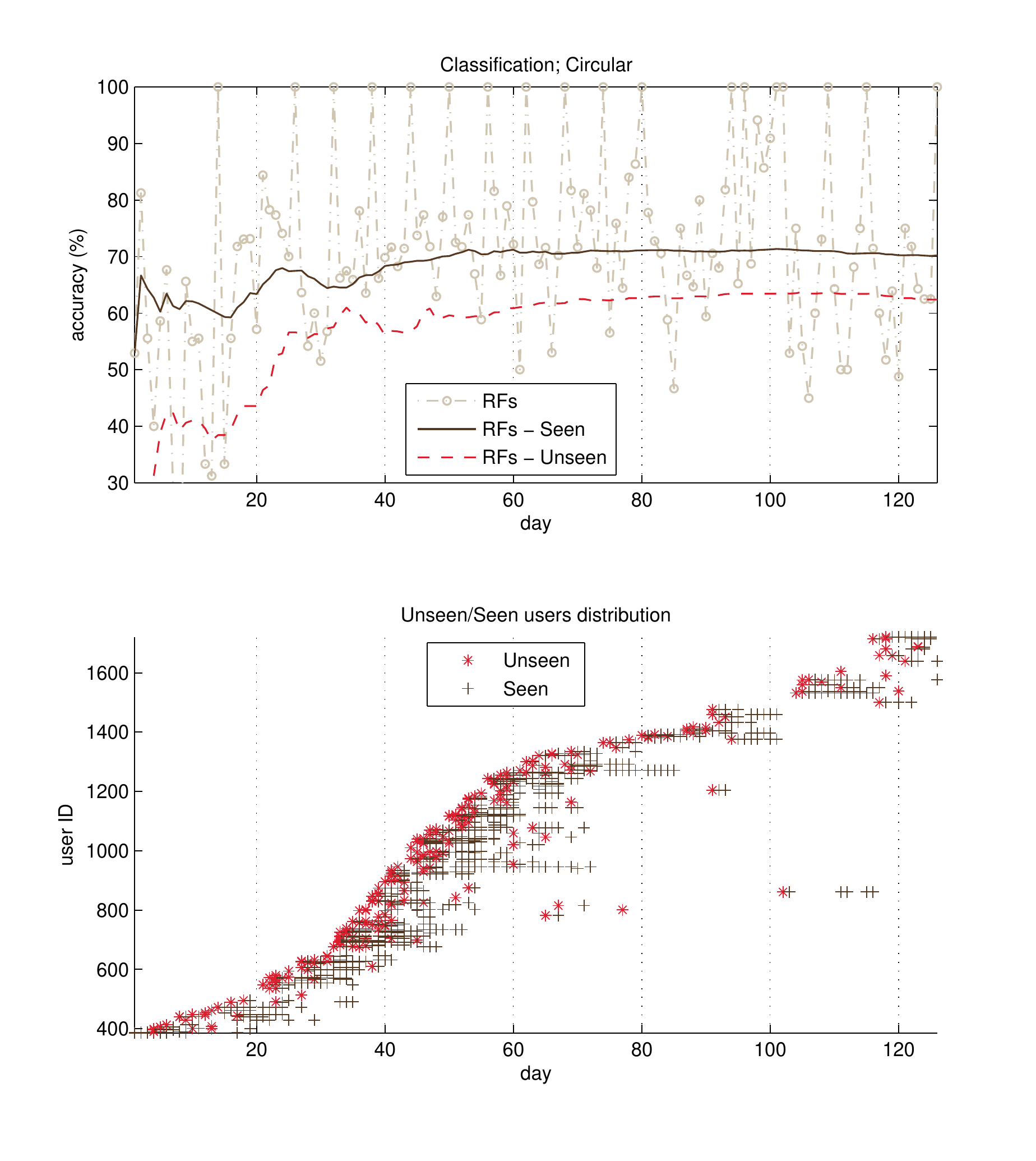, width=0.99\linewidth}
 \caption{Test accuracy performance along with arrival of sequential data. The incoming unseen activity data is predicted based on learned model using previous training data to obtain the test accuracy. First day test is conducted when a model is learned with 3 training days data. }
 \label{fig:testdist}
\end{figure}

In Figure \ref{fig:testdist}, we plot the test accuracy performance along with arrival of sequential data. The incoming unseen activity data is predicted based on learned model using previous training data to obtain the test accuracy. Subsequently, this tested data is used for training in next sequence day based on its true (labelled by users) activity label. A test data is coming either from unseen user or seen user. Seen user means that his/her activity history is used during training models, and unseen user is not. As shown in the bottom figure in Figure \ref{fig:testdist}, unseen users are appearing almost every days from multiple users. The top figure in Figure \ref{fig:testdist} shows accumulative accuracy of RF WMV where the values are averaged for seen users (solid line) and unseen users (dashed line) respectively.  By accumulative accuracy, we mean the average accuracy of the system from test day 1 to the current test day. We see that the classification accuracy for seen users are better than unseen users which shows that learning from users' own history helps to improve the classification accuracy. Classification performance of unseen users improves as the training day accumulates more than that of seen users. For test classification of unseen user, the model learned from cross-user and users from social-demographics are used. Since there are more number of training data from cross-user and social demographics based users than user-specific information, the performance could be improved relatively larger than that of seen user case.   

\begin{figure}[htb]
 \centering
   \begin{tabular}{c c  }
  \hspace{-0.17in}   \subfigure[Prediction]{\includegraphics[width=1.8in]{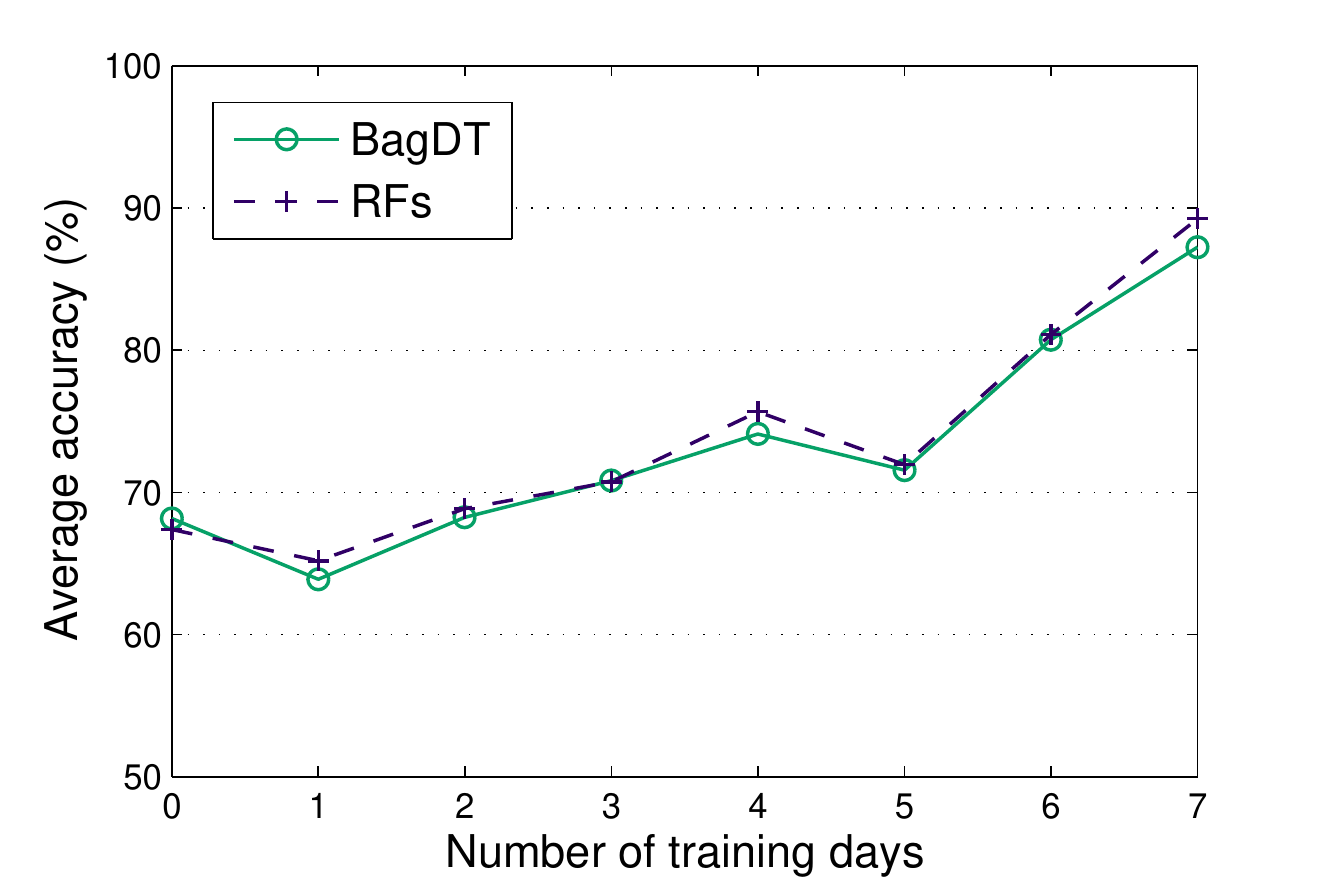} } &
 \hspace{-0.4in}   \subfigure[Number of users and cases]{\includegraphics[width=1.8in]{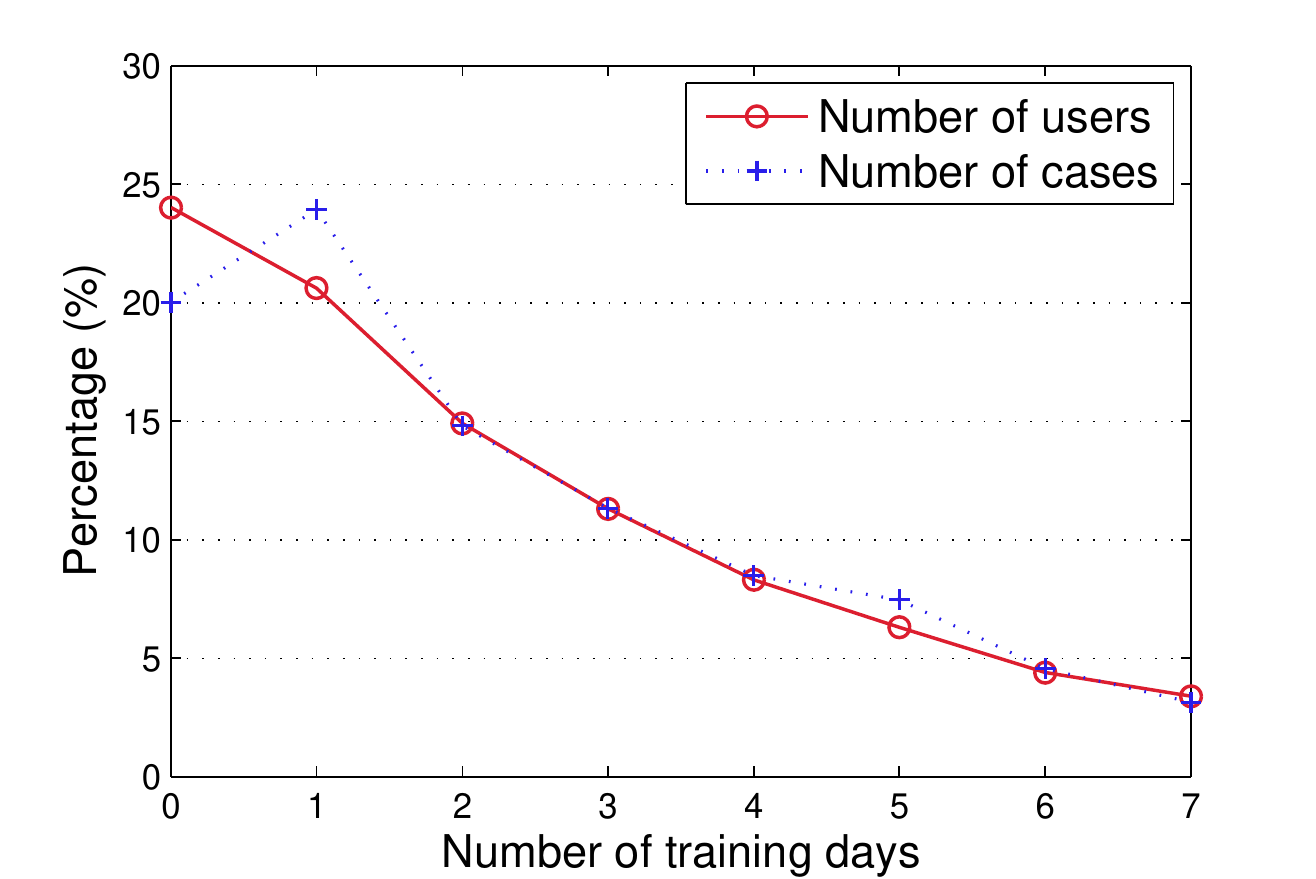}} 
 \end{tabular}
 \caption{(a) Averaged accuracy along with the number of user-specific training days for individual users. (b) Corresponding number of users and test cases during testing}
\label{fig:daypredict}
\end{figure}

To observe the effect of number of user-specific training days further, average classification accuracy is shown along with user-specific training days again. Different from settings in Figure \ref{fig:trainingdays}, every user has different total number of training days for learning in Figure \ref{fig:daypredict}. Training days `0' indicates that no user-specific data is used in training for that user (unseen user). In Figure \ref{fig:daypredict}, an average accuracy value increases as number of user-specific training days increases. To avoid a biased result, test results involving more than 30 users at that day are shown. Decay value at day 1 is related to bias effect from small individual user sample size. A reason of decay at training day 5 in Figure \ref{fig:daypredict} (a) may be found from that the number of test cases are relatively more than the number of the users. Ratio (number of test samples versus number of test users) at training days 5 (including day 1) is relatively higher than other cases\footnote{Ratio at day 5 is 5.92 and 5.8 at day 1. Average of others [0,2,3,4,6,7] days is 4.84. Ratio = [4.1667,    5.8010,    4.9597,    5.0088,    5.1084,  5.9206,    5.1818,    4.6176].}.
It means that each user has more activity points than other cases in average, so more unseen/unusual activity patterns would be included in that day 5 case than other cases.

Most of users have less than 3 training days as shown in Figure \ref{fig:daypredict} (b). If more individual users have more training days, overall accuracy of seen user (in Figure \ref{fig:testdist}) could be improved. We can observe that average accuracy keep improves as training days increases in Figure \ref{fig:daypredict} (a).

%
%
%
%
\section{Conclusions}
In this paper, we proposed a framework to recognize an activity type of a traveler when his/her movement is tracked by mobile sensors, as per our Future Mobility Survey (FMS) technology \cite{Cottrill2013FMS}. With different shapes of spatial quantization, ensemble classifiers are applied to process noisy real-world spatial-temporal and contextual data.
To improve generalization performance, our model takes advantage of cross-user historical data as well as user-specific information, including social demographic characteristics. Fusion of multiple classifiers learned from different user populations shows improved generalization performance than that of individual classifier learning. We evaluated the activity classification performance along with sequential data for a real life situation. As the number of training data is accumulating, the generalization performance is improved. Also, we demonstrated that learning from a user's own history improves the recognition accuracy. Our empirical results demonstrate that the proposed method contributes significantly to our travel survey application.

In terms of future work, there are several potential avenues for investigation. To find the centroids of Voronoi polygon, more adaptive spatial clustering techniques such as hierarchical clustering and density based clustering could be used \cite{Liu2010mobilityclustering,Kisilevich2010spatialtemporalclsutering}. We can compare between pointwise classification (deployed in the current system) and sequence based classification (HMM, CRF, etc.) which is workable for continuous travel data environment. Finally, we can assess the positive feedback cycle between the algorithm and user labeling to improve classification performance in future survey.



\section{Acknowledgments}
This research was funded by the Singapore National Research Foundation (NRF) through the Singapore-MIT Alliance for Research and Technology (SMART) Center for Future Urban Mobility (FM) Group. The authors would like to thank FMS team (Bruno, In\^{e}s, Kalan, Rui) for their support and give special thanks to Rudi Ball and Carlos Carrion for their data cleaning effort.

%
\bibliographystyle{abbrv}
\bibliography{ActivityRecognition_KDD}

%
%


\end{document}